\documentclass[twocolumn,epjc3]{svjour3}          
\pdfoutput=1

\RequirePackage[T1]{fontenc}

\smartqed  

\RequirePackage{footmisc}
\RequirePackage{textcomp}
\RequirePackage{graphicx}
\usepackage{feynmf}
\usepackage{amsmath}
\usepackage{upgreek}
\RequirePackage{mathptmx}      
\RequirePackage[numbers,sort&compress]{natbib}
\usepackage{textpos}
\usepackage{lineno}
\RequirePackage{hyperref}
\usepackage{ragged2e}
\graphicspath{ {./figures/} }
\usepackage{float}
\usepackage{everysel}
\journalname{Eur. Phys. J. C}

\begin{document}  
\title{Physics potential for the measurement of  ${\sigma(H\nu\bar{\nu})\times \text{BR}(H\rightarrow\mu^+\mu^-)}$ at the 1.4 TeV CLIC collider\thanksref{t1}}

\author{G. Milutinovi\'{c}-Dumbelovi\'{c}\thanksref{e1,addr1}
        \and
        I. Bo\v{z}ovi\'{c}-Jelisav\u{c}i\'{c}\thanksref{addr1} 
        \and
        C. Grefe\thanksref{addr2, addr3}
        \and
        G. Ka\u{c}arevi\'{c}\thanksref{addr1} 
        \and \\
        S. Luki\'{c}\thanksref{addr1}
        \and
        M. Pandurovi\'{c}\thanksref{addr1}
        \and
        P. Roloff\thanksref{addr3} 
        \and
        I. Smiljani\'{c}\thanksref{addr1} 
}

\thankstext[$\star$]{t1}{This work was carried out in the framework of the CLICdp collaboration}
\thankstext{e1}{e-mail: gordanamd@vinca.rs}

\institute{Vinca Institute of Nuclear Sciences, University of Belgrade, Mihajla Petrovi\'{c}a Alasa 12-14, 11001 Belgrade, Serbia\label{addr1}
          \and
          Universit\"{a}t Bonn, D-53012 Bonn, Germany\label{addr2}
          \and
          CERN, CH-1211 Geneva 23, Switzerland\label{addr3}
}

\date{Received: date / Accepted: date}

\setlength{\TPHorizModule}{1cm}
\setlength{\TPVertModule}{1cm}

\maketitle
\begin{abstract}

The future Compact Linear Collider (CLIC) offers a possibility for a rich
precision physics programme, in particular in the Higgs\thinspace sector\thinspace through\thinspace the\thinspace
energy\thinspace staging.\thinspace This is the\thinspace first\thinspace paper\thinspace addressing\thinspace the\thinspace measurement\thinspace
of\thinspace the Standard Model Higgs boson decay\thinspace
into\thinspace two\thinspace muons\thinspace at\thinspace 1.4\thinspace TeV CLIC. With
respect to similar studies at future linear colliders, this paper
includes several novel contributions to the statistical
uncertainty of the measurement. The later includes the Equivalent Photon
Approximation and realistic forward electron tagging based on energy
deposition maps in the forward calorimeters, as well as several processes
with the Beamstrahlung photons that results in irreducible contribution to
the signal. In addition, coincidence of the Bhabha scattering with the
signal and background processes is considered, altering the signal
selection efficiency. The study is performed using a fully simulated
CLIC\_ILD detector model. It is shown that the branching ratio for the
Higgs decay into a pair of muons BR(${H\rightarrow\mu^+\mu^-}$) times the Higgs
production cross-section in $WW$-fusion $\sigma(H\nu\bar{\nu})$ can be measured with 38\%
statistical accuracy at ${\sqrt{s} =\text{1.4 TeV}}$, assuming an integrated luminosity of
1.5 ab$^{-1}$ with unpolarised beams. If 80\% electron beam polarisation is
considered, the statistical uncertainty of the measurement is reduced to
25\%. Systematic uncertainties are negligible in comparison to the
statistical uncertainty.

\end{abstract}

\section{Introduction}

\justify
Measurements of Higgs branching ratios, and consequently Higgs couplings, provide a strong test of the
Standard Model (SM) and possible physics beyond. Models that could possibly extend the SM Higgs sector
(Two Higgs Doublet model, Little Higgs models or Compositeness models) will require Higgs
couplings to electroweak bosons and Higgs-fermion Yukawa couplings (coupling-mass linearity) to deviate from the
SM predictions \cite{r1,r2}. 

The Compact Linear Collider (CLIC) represents an excellent environment to study properties of the
Higgs boson, including its couplings, with a very high precision \cite{r3,r4}. Measurements of rare ${H\rightarrow\mu^+\mu^-}$ decays 
are particularly challenging because of the very low branching ratio of 2${\times 10^{-4}}$ predicted in the SM \cite{r5} for a Higgs mass of \/ \/126\thinspace GeV. 
Current results indicate that the LHC was not able to access Higgs coupling to muons ($g_{H{\mu\mu}}$), based on the runs at 7 TeV and 8 TeV centre-of-mass (CM) energies \cite{r6}.   
Projections for the HL-LHC, assuming 300 fb${^{-1}}$ and 3 ab${^{-1}}$ of data, predict uncertainties of 23\% and 8\% respectively for the $g_{H{\mu\mu}}$ coupling \cite{r7}.
In order to provide the best physics reach in the shortest time and for an optimal cost, the operation of the CLIC accelerator is foreseen in energy stages of 350\thinspace GeV,
1.4 TeV\thinspace and\thinspace 3\thinspace TeV\thinspace \cite{r8}. At 1.4 TeV and 3 TeV, sufficiently large Higgs boson samples can be produced to allow studies of rare Higgs decays\thinspace. 
A sample of 3.7 ${\times 10^{5}}$ Higgs bosons\thinspace can be\thinspace produced at 1.4 TeV\thinspace CM\thinspace energy,\thinspace 
for an integrated luminosity of 1.5 ab${^{-1}}$ with unpolarised beams. With the expected instantaneous luminosity of 3.2${\times 10^{34}}$cm${^{-2}}$s${^{-1}}$ this can be achieved in approximately five years of detector operation, 
with 200 running days per year and an effective up-time of 50\% \cite{r9}.
The signal\thinspace sample\thinspace size will\thinspace be\thinspace doubled at 3 TeV CM energy due to rising cross-section for $WW$-fusion \cite{r10}. 

A similar study has been performed at 3 TeV CM energy \cite{r10}. Compared to 
the study at 3 TeV, several challenges for the measurement of $H\rightarrow\mu^+\mu^-$ at 
CLIC are discussed for the first time in this paper. First, background 
processes with photons in the initial state simulated using both the 
expected Beamstrahlung spectrum at CLIC and the Equivalent Photon 
Approximation (EPA) \cite{r11, r12}, were considered. Forward electron tagging (Section \ref{section5}) 
leads to a rejection of 48\% and 42\% of the $e^-e^+\rightarrow e^-e^+\mu^+\mu^-$ and ${e^\pm}\gamma\rightarrow{e^\pm}\mu^+\mu^-$ background events, respectively. 
The impact of Bhabha scattering events on the rejection of events with forward electrons is investigated.  

The paper is organised as follows. In Section \ref{section2} the simulation tools used for the analysis
are listed and in Section \ref{section3} the CLIC\_ILD detector model is briefly described. Signal and background
processes and event samples are discussed in Section \ref{section4}. Tagging of background high-energy electrons is described in Section \ref{section5}. Event preselection
and the final selection based on a Multivariate Analysis
(MVA) approach are described in Section \ref{section6}. The di-muon invariant mass fit and the extraction of the statistical
uncertainty of the measurement are described in Section \ref{section7}. In Section \ref{section8} the impact of electron polarisation on the statistical 
uncertainty of the ${\sigma(H\nu\bar{\nu})\times \text{BR}(H\rightarrow\mu^+\mu^-)}$ measurement is described.
Systematic uncertainties are discussed in Section \ref{section9}, followed by the conclusions in Section \ref{section11}. 

\section{Simulation and analysis tools}
\label{section2}
  
Higgs production through $WW$-fusion is
simulated in Whizard 1.95 \cite{r13,r14} including a realistic CLIC
beam spectrum and initial state radiation. The
generator Pythia 6.4\thinspace\cite{r15} is used to simulate the
Higgs decay into two muons. The CLIC luminosity spectrum and beam-induced processes are obtained by GuineaPig 1.4.4 \cite{r16}. Background events
are also generated with Whizard using Pythia 6.4
to simulate the hadronization and fragmentation
processes. Simulation of tau decays is done by Tauola \cite{r17}. 
The CLIC\_ILD detector simulation is
performed using Mokka \cite{r18} based on Geant4 \cite{r19}. Before digitisation of
the detector signals, pile-up from $\gamma\gamma\rightarrow \text{hadrons}$ interactions is overlaid on the physics events.
The particle flow algorithm, PandoraPFA \cite{r20,r21} is employed in 
reconstruction of the final-state particles within the Marlin reconstruction framework \cite{r22}. The TMVA
package \cite{r23} is used to separate signal from
background by MVA of signal
and background kinematic properties.

\section{The CLIC\_ILD detector model}\label{section3}
\justify
The ILD detector concept \cite{r24} is modified for CLIC
according to the specific experimental conditions at higher energies\thinspace\cite{r3}. 
The subsystems of particular relevance for the presented analysis are discussed here. A complete description of the CLIC\_ILD detector can be found in \cite{r25} . 

The main tracking device of CLIC\_ILD is the Time Projection Chamber (TPC) providing a point resolution in the $r\phi$ plane better than 100 ${\upmu}$m,
for charged particles in the detector angular acceptance \cite{r3}. Additional silicon trackers cover polar angles down to 7\textdegree. They 
have a single point resolution of 7 ${\upmu}$m, and together with the TPC 
improve the tracking accuracy in the $r\phi$ plane. In order to provide precision tracking and vertexing closer to the beam-pipe, 
a Vertex Detector capable of an impact parameter resolution of 3 ${\upmu}$m \cite{r26} is foreseen. Calorimetry at CLIC is 
based on fine-grained sandwich calorimeters optimized for particle-flow analysis (PFA). PFA is based on reconstruction of 
four-vectors of visible particles, combining the information from precise tracking with highly granular calorimetry. 
The detector comprises a central solenoid magnet, with\thinspace a\thinspace field\thinspace of\thinspace 4\thinspace T. High\thinspace muon\thinspace reconstruction\thinspace efficiency of 99\%, for muons above 7.5 GeV, 
is achieved by combining information from the central tracker (TPC plus silicon tracker) with information provided by the iron yoke instrumented 
with the 9 layers of Resistive Plate Chamber detectors.

In principle, hadrons produced in the interaction of the beam-induced photons affect the TPC occupancy and consequently the muon reconstruction efficiency. However, 
in the studied sample of muons from ${H\rightarrow\mu^+\mu^-}$ decays, muon reconstruction efficiency is above 99\% in the barrel region, in the presence of 
$\gamma\gamma\rightarrow \text{hadrons}$.

The average muon transverse momentum resolution for the signal sample is $\Delta(1/p_\text{T}) = 3.3\times10^{-5} \text{GeV}^{-1}$ in the barrel region.
The impact of transverse momentum resolution on\thinspace the\thinspace statistical\thinspace uncertainty of ${\sigma(H\nu\bar{\nu})\times\text{BR}(H\rightarrow\mu^+\mu^-)}$ measurement is discussed in Section \ref{section9}. 

In the very forward region of the CLIC\_ILD detector, below $\theta$=8\textdegree, no tracking information is available. The region between 0.6\textdegree \/ and 6.3\textdegree \/ is instrumented with the two silicon-tungsten sampling calorimeters, LumiCal and BeamCal \cite{r27}, for the luminosity measurement, beam-parameter control,
\/ as well as for the tagging of high-energy electrons escaping the main detector at low angles.
Together with the very forward segments of the electromagnetic calorimeter covering the polar-angle region between 6.3\textdegree and 8\textdegree, 
it is possible to suppress the four-fermion SM background with the characteristic low-angle electron signature. 
The simulation of the very-forward electron tagging is described in Section \ref{section5}. 

\section{Event samples}\label{section4}

\begin{figure}[H]
\centering
\caption{\label{fig-WWfusion} Feynman diagram of the Higgs production in $WW$-fusion and the subsequent Higgs boson decay to a pair of muons.}
\includegraphics[width=\columnwidth, clip, trim = 2.5cm 19.cm 5cm 3.8cm]{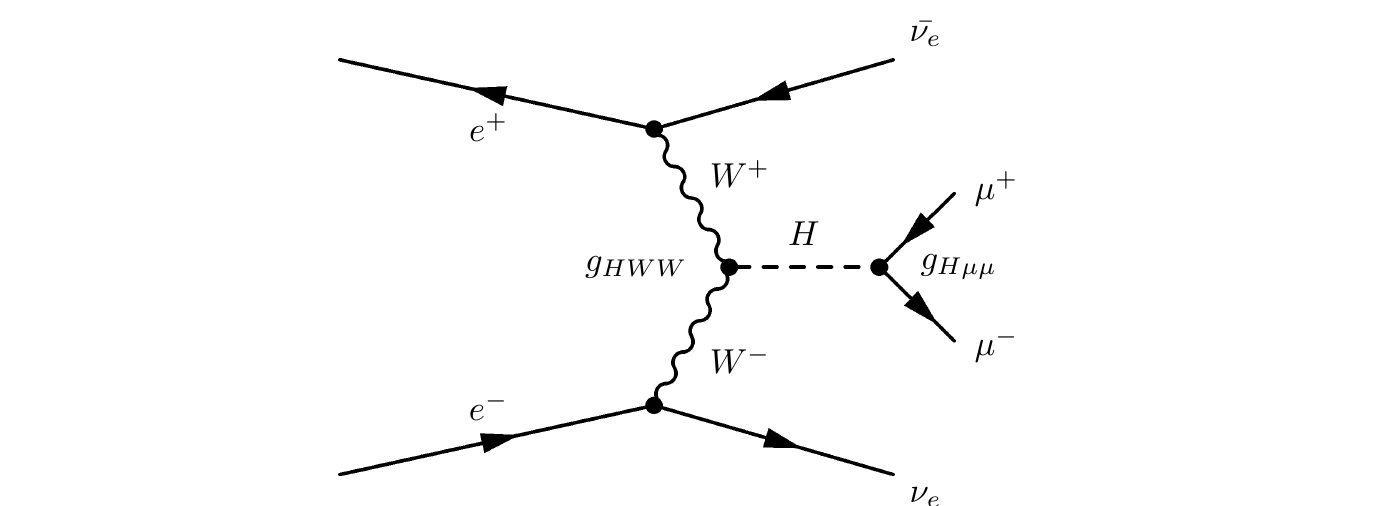}
\end{figure}

At ${\sqrt{s}=1.4}$\thinspace TeV the SM Higgs boson is predominantly produced via $WW$-fusion (Figure \ref{fig-WWfusion}). 
The effective cross-section for Higgs production in $WW$-fusion is 244 fb without beam polarization. 
The Higgs production cross-section above 1 TeV can be measured with a statistical precision better than 1\% as shown in \cite{r1}. 
The $e^+e^-\rightarrow H\nu\bar{\nu}, H\rightarrow\mu^+\mu^-$ signal statistics are expected to be small (of the order of a few tens of events) because 
of the small branching fraction for this particular decay. 

We have simulated a sample of 24000 signal events, rou- ghly corresponding to 300 times the number of events expected in 1.5 ab${^{-1}}$ of data. 
This is needed in order to provide an adequate description of the signal Probability Density Function (PDF) (Section \ref{section71}). 
The signal and the dominant background processes are listed in Table \ref{tableprocess}. For each of the background processes, 
samples of 2 ab${^{-1}}$  are generated.

In addition to the processes listed in Table \ref{tableprocess}, we\thinspace have considered s-channel\thinspace $e^+e^-\rightarrow\mu^+\mu^-$ production,\thinspace as\thinspace well\thinspace as several\thinspace
processes\thinspace with\thinspace tau\thinspace pair\thinspace in\thinspace the\thinspace final\thinspace state $e^+e^-\rightarrow \tau^+\tau^-$, $e^+e^-\rightarrow \nu_{\tau}\bar{\nu_{\tau}}\tau^+\tau^-$, $e^+e^-\rightarrow e^-e^+\tau^+\tau^-$. 
Tau decays become relevant if both taus decay into two muons which happens in $\sim$3\% of cases \cite{r28}. However, the
invariant mass of the di-muon system will not match the Higgs mass window considered in this analysis (see Section \ref{section61}). 
The same holds for $e^+e^-\rightarrow\mu^+\mu^-$ production.

In addition, processes with quasi-real photons in the initial state 
are simulated using the EPA. In this analysis, such events are grouped together with the processes involving Beamstrahlung photons.
In this way, processes with roughly similar\thinspace kinematic\thinspace characteristics\thinspace are\thinspace grouped\thinspace together\thinspace, as sh-own in Table \ref{tableprocess}. 
The notation ${e^\pm}\gamma$ represents the sum of cross-sections for the processes with either ${e^-}\gamma$ or ${e^+}\gamma$ in the inital state.

At ${\sqrt{s} =1.4}$ TeV, the Higgs boson is also produced via $ZZ$-fusion, with a cross-section of about 10\% of the Higgs production cross-section in $WW$-fusion. 
However, on a test sample of 300 $ZZ$-fusion events followed by the Higgs decay to a pair of muons, not a single event passed the selection described in Sections 
\ref{section61} and \ref{section62}. This implies an efficiency smaller than 1.2\% (95\% CL) for this channel equivalent to
less than 0.1 events passing the final selection. Therefore, the Higgs production through $ZZ$-fusion is not considered relevant for this analysis.

Photons, dominantly emitted by Beamstrahlung, produce incoherent 
pairs deposited mainly in the low-angle calorimeters. On average, 1.3 two-photon interactions producing had-ronic final states 
occur per bunch crossing \cite{r29} which may affect the muon reconstruction 
in the tracking detectors. These hadrons are included in the analysis by overlaying 60 bunch crossings in the simulation, before the 
digitisation and event reconstruction phase. These events, as well as other physics events, are passed trough the full detector simulation \cite{r30}. 

\begin{table}[h] 
\centering
\caption{\label{tableprocess} List of considered processes with corresponding cross-sections. Cross-section values marked by * are generated with the additional kinematic requirements: 100 GeV< $m_{\mu\mu}<$150 GeV, and 8\textdegree<$\theta_{\mu}$<172\textdegree, where $m_{\mu\mu}$ stands for di-muon invariant mass and $\theta_{\mu}$ is the polar angle of the reconstructed muon. The cross-sections for all processes with photons in the initial state include both Beamstrahlung and processes with EPA photons. Cross-sections for processes ${e^\pm}\gamma\rightarrow{e^\pm}\mu^+\mu^-$ and ${e^\pm}\gamma\rightarrow{e^\pm}\nu_{\mu}\bar{\nu}_{\mu}\mu^+\mu^-$ represent the sum of cross-sections for the processes with both initial states ${e^-}\gamma$ and ${e^+}\gamma$.}
\begin{tabular*}{\columnwidth}{@{\extracolsep{\fill}}ll@{}}
\hline
\multicolumn{1}{@{}l}{Process}  & $\sigma(fb)$ \\
\hline
$e^+e^-\rightarrow H\nu\bar{\nu}, H\rightarrow\mu^+\mu^-$              & 0.0522         \\
$e^+e^-\rightarrow\nu\bar{\nu}\mu^+\mu^-$                      & 129         \\
$e^-e^+\rightarrow e^-e^+\mu^+\mu^-$                 & 24.5${^*}$         \\
${e^\pm}\gamma\rightarrow{e^\pm}\mu^+\mu^-$                       & 1098${^*}$        \\
${e^\pm}\gamma\rightarrow{e^\pm}\nu_{\mu}\bar{\nu}_{\mu}\mu^+\mu^-$                           & 30          \\
$\gamma\gamma\rightarrow\nu_{\mu}\bar{\nu}_{\mu}\mu^+\mu^-$                     & 162          \\
$e^+e^-\rightarrow{e^+e^-}\nu_{\mu}\bar{\nu}_{\mu}\mu^+\mu^-$                         & 1.6           \\
\hline
\end{tabular*}
\end{table}

\section{Tagging of EM showers in the very forward region}\label{section5}

In the polar angle region below $\theta$=8\textdegree, tracking information and hadronic calorimetry are not available. The four-fermion background 
$e^+e^-\rightarrow e^+e^-\mu^-\mu^+$ of multiperipheral type and similar processes like ${e^\pm}\gamma\rightarrow{e^\pm}\mu^-\mu^+$ can fake the missing energy signature of the signal 
if the final state electrons (spectators), emitted at the polar angles smaller than $\theta$=8\textdegree, escape undetected. 

Electron detection in the very forward region involves the reconstruction of electromagnetic showers in the presence of intense beam-induced background depositing in the very forward 
calorimeters a large number of low-energy particles, mostly incoherent pairs from Beamstrahlung \cite{r31}. This deposition amounts to several hundred thousand of $e^+e^-$ pairs per bunch crossing \cite{r32}. 

Furthermore, Bhabha events where one or both electrons are detected in the very forward calorimeters may occur in coincidence with either signal or background, 
even within the 10 ns time stamp. Tagging of such Bhabha electrons will result in the rejection of signal (or background). In order to prevent significant loss 
of signal statistics, the electron tagging was optimized to identify showers with energy higher than 200 GeV and a polar angle above 
1.7\textdegree \thinspace only\footnote{Simulation with the reconstruction algorithm from Ref.\cite{r33} shows that assuming these cuts, the reconstruction efficiency, in BeamCal and LumiCal is above 98\%, with a negligible fake rate.}. 
Under these requirements, the loss of the number of signal events due to tagging of Bhabha electrons amounts to 7\%. Out of these 7\%, in slightly more than a half of events one
electron is added to the final state and, in the remainder two Bhabha electrons are added. Table \ref{tab-tagging} shows rejection rates for signal and background obtained by the forward electron\thinspace tagging\thinspace due\thinspace to\thinspace Bhabha pile-up.

In conclusion, very forward tagging of high-energy electrons serves to half the fraction of background with spectator electrons, 
with a moderate loss of signal of 7\% in the presence of Bhabha coincidence. 

\begin{table}[hb!]
\centering
\caption{\label{tab-tagging} Rejection rates for signal and background by the forward electron tagging.}
\begin{tabular*}{\columnwidth}{@{\extracolsep{\fill}}ll@{}}
\hline
\multicolumn{1}{@{}l}{Process} &\parbox{2.5cm}{\vspace{.2\baselineskip}Rejection rate \vspace{.2\baselineskip}}\\
\hline
$e^-e^+\rightarrow e^-e^+\mu^+\mu^-$                  &  48\%   \\
${e^\pm}\gamma\rightarrow{e^\pm}\mu^+\mu^-$       & 42\%  \\                 
Signal                             & 7\% \\                           
\hline
\end{tabular*}
\end{table}

\section{Event selection}\label{section6}

The event selection is done in two steps. First, a preselection is performed aiming to suppress background originating 
from beamstrahlung as well as the processes with spectator electrons described in Section \ref{section5}. The final event selection 
uses a multivariate classifier based on boosted decision trees (BDT) to suppress remain background processes on the basis of their kinematic properties.

\subsection{Preselection}\label{section61}

In order to suppress the impact of the beam-induced background, only reconstructed particles with transverse momenta $p_\text{T}$ > 5 GeV are used in the analysis. 
Furthermore, the preselection of events was made by requiring a reconstruction of exactly two muons in the event, with an invariant mass of the di-muon system in the 
window centered around the Higgs mass 105-145 GeV. In addition, the absence of tagged electrons 
with energy above 200 GeV and polar angle above 1.7\textdegree \thinspace \thinspace is required in order to suppress background with spectator electons emitted in the very forward region 
of the detector.

\subsection{MVA selection}\label{section62}

As a second step in the event selection, MVA techniques are used based on the BDT classifier implemented in the TMVA package. 
From the signal sample, quater of all events are reserved for TMVA training, 
as well as 0.5 ab${^{-1}}$ of each background. The following observables were used for the classification\thinspace of\thinspace events\thinspace, similar\thinspace to\thinspace the\thinspace CLIC\thinspace study\thinspace at\thinspace ${\sqrt{s} =3}$ TeV \cite{r10}:

\begin{itemize}
   \item visible energy of the event excluding the energy of the di-muon system, $E_\text{vis}$,
   \item transverse momentum of the di-muon system, $p_\text{T}(\mu\mu)$,
   \item scalar sum of the transverse momenta of the two selected muons, 
         $p_\text{T}(\mu_1)+ p_\text{T}(\mu_2)$,
   \item boost of the di-muon system, $\beta_{\mu\mu} = \left | p_{\mu\mu} \right | / E_{\mu\mu}$,
   \item polar angle of the di-muon system, $\theta_{\mu\mu}$,
   \item cosine of the helicity angle, 
         $\cos\theta^*$. 
\end{itemize}

\begin{figure}[H]
\centering
\caption{\label{eVis}Distribution of the visible energy for the signal and \/ ${e^\pm}\gamma\rightarrow{e^\pm}\mu^+\mu^-$, $e^+e^-\rightarrow\nu\bar{\nu}\mu^+\mu^-$, ${e^\pm}\gamma\rightarrow{e^\pm}\nu_{\mu}\bar{\nu}_{\mu}\mu^+\mu^-$ and $e^+e^-\rightarrow e^+e^-\mu^+\mu^-$ background. }
\includegraphics[width=\columnwidth]{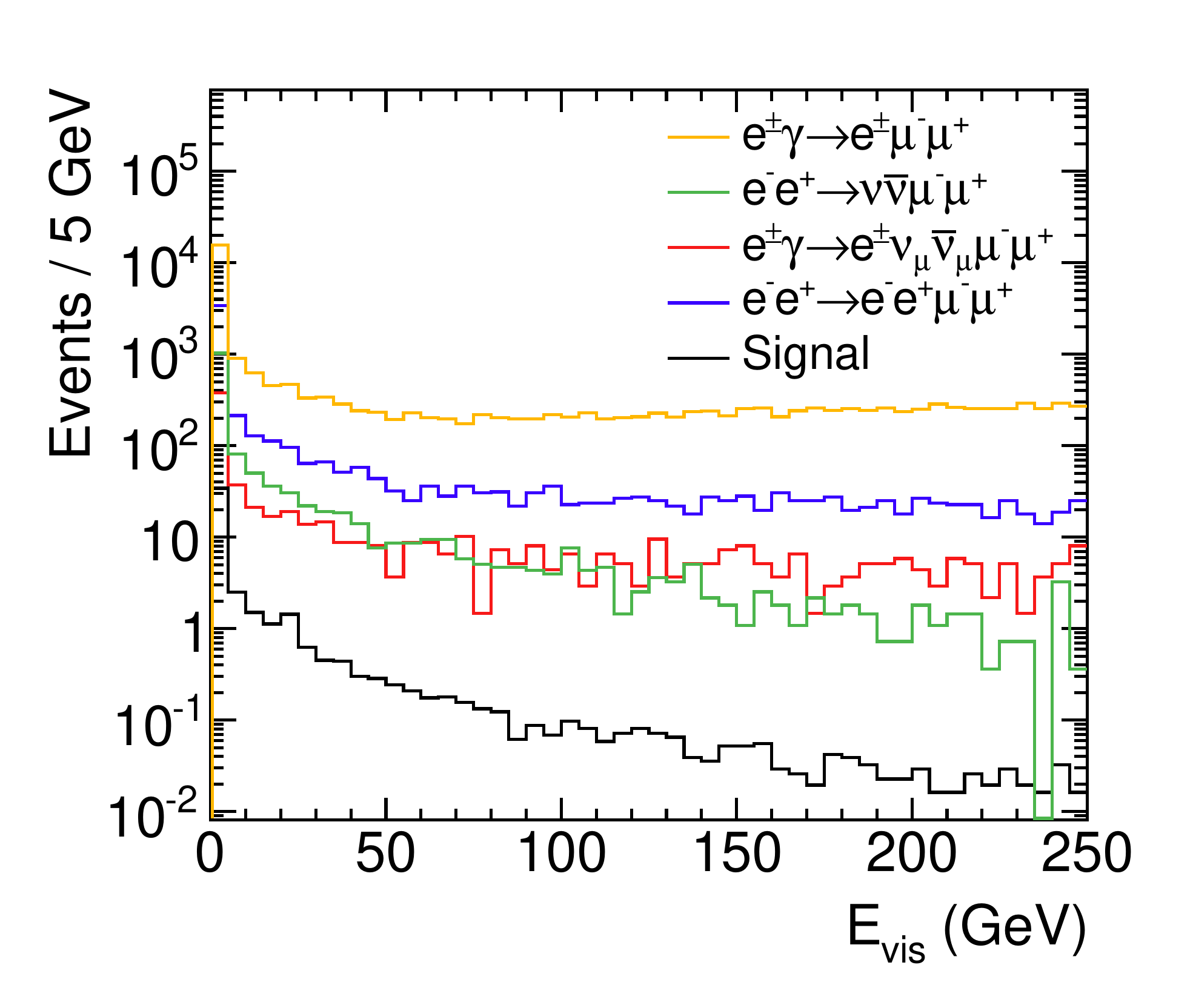}
\begin{textblock}{1.}(2.,-6.)
\bf CLICdp
\end{textblock}
\end{figure}

\begin{figure}[H]
\centering
\caption{\label{fig-pt}Distribution of the $p_\text{T}(\mu\mu)$ for the signal and \/ $e^+e^-\rightarrow\nu\bar{\nu}\mu^+\mu^-$,  $\gamma\gamma\rightarrow\nu_{\mu}\bar{\nu}_{\mu}\mu^+\mu^-$, \/ $e^-e^+\rightarrow e^+e^-\mu^+\mu^-$ background with spectator electrons. }
\includegraphics[width=\columnwidth]{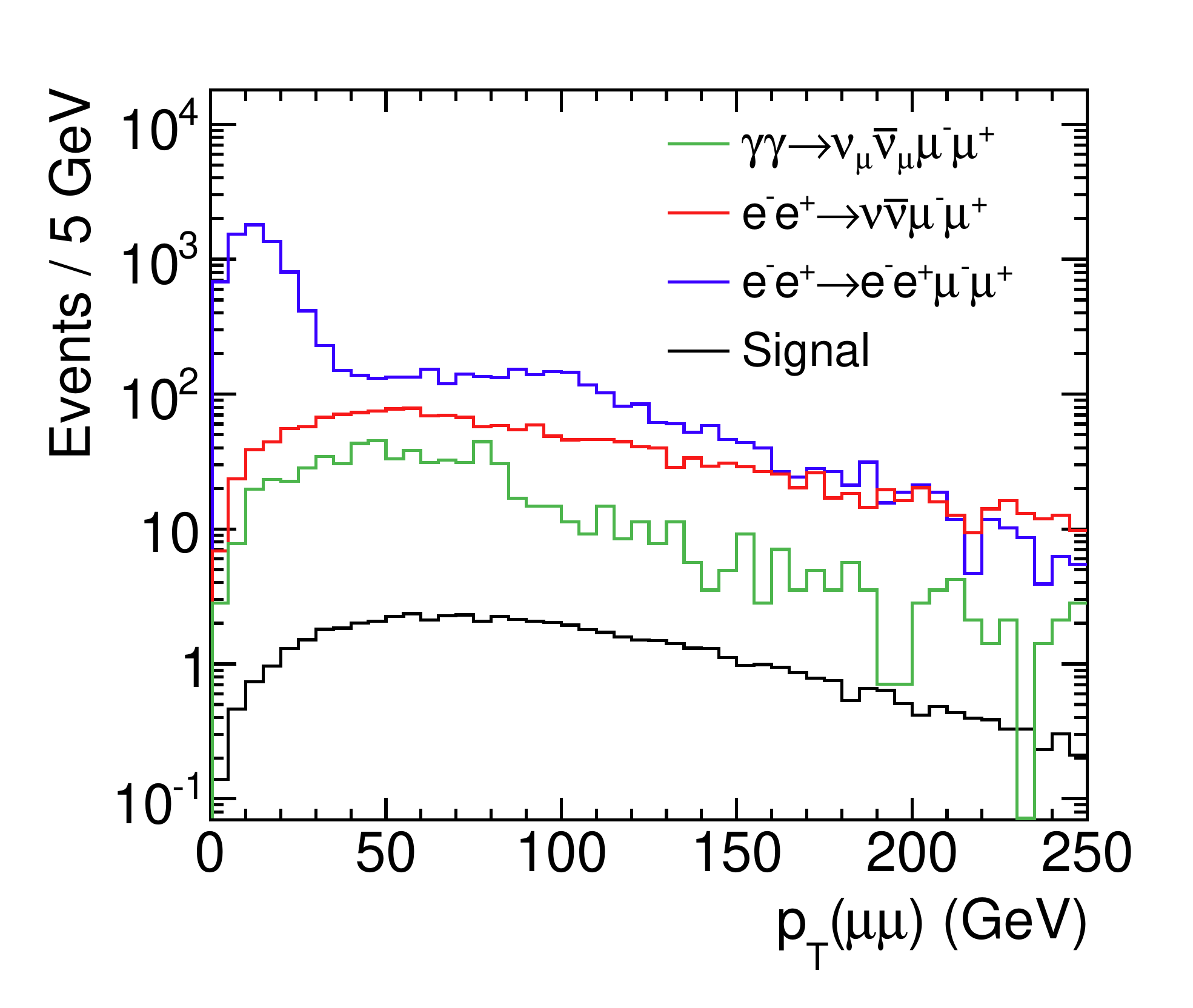}
\begin{textblock}{1.}(2.2,-6.1)
\bf CLICdp
\end{textblock}
\end{figure}

The process $e^+e^-\rightarrow\nu\bar{\nu}\mu^+\mu^-$, with the same final state as the signal, represents an irreducible background 
and can not be substantially suppressed before the invariant mass fit of the di-muon system. 
The process $\gamma\gamma\rightarrow\nu_{\mu}\bar{\nu}_{\mu}\mu^+\mu^-$ has a similar final state, 
but a different CM energy distribution in the initial state, since it involves Beamstrahlung or EPA photons rather than initial electrons. 
This leads to a different distribution of the boost of the di-muon system, allowing separation from the signal to some extent. 
The processes $e^+e^-\rightarrow\nu\bar{\nu}\mu^+\mu^-$ and $\gamma\gamma\rightarrow\nu_{\mu}\bar{\nu}_{\mu}\mu^+\mu^-$, 
have slightly different distributions of the helicity angle from the signal. All processes with one or two spectator electrons show significant differences from the signal, 
primarily in the distribution of the visible energy (Figure \ref{eVis}). These processes are also effectively suppressed by the $p_\text{T}(\mu_1) + p_\text{T}(\mu_2)$ observable. 
In addition, for the $e^+e^-\rightarrow e^+e^-\mu^+\mu^-$  process, the distribution of $p_\text{T}(\mu\mu)$ exhibits a peak at lower values than the signal 
(Figure \ref{fig-pt}). This peak corresponds to events in which the di-muon system recoils against electron spectators or outgoing photons that are emitted
below the angular cut of the very forward EM-shower tagging. The above is illustrated in Figure \ref{fig-pt} showing the $p_\text{T}$ distributions for representative 
background processes.

\begin{figure*} 
\centering
\caption{\label{fig-BDT}Stacked histograms of the BDT output variable for signal and background processes (a); Significance as a function of the BDT cut value (b).}
\includegraphics[width=.57\textwidth]{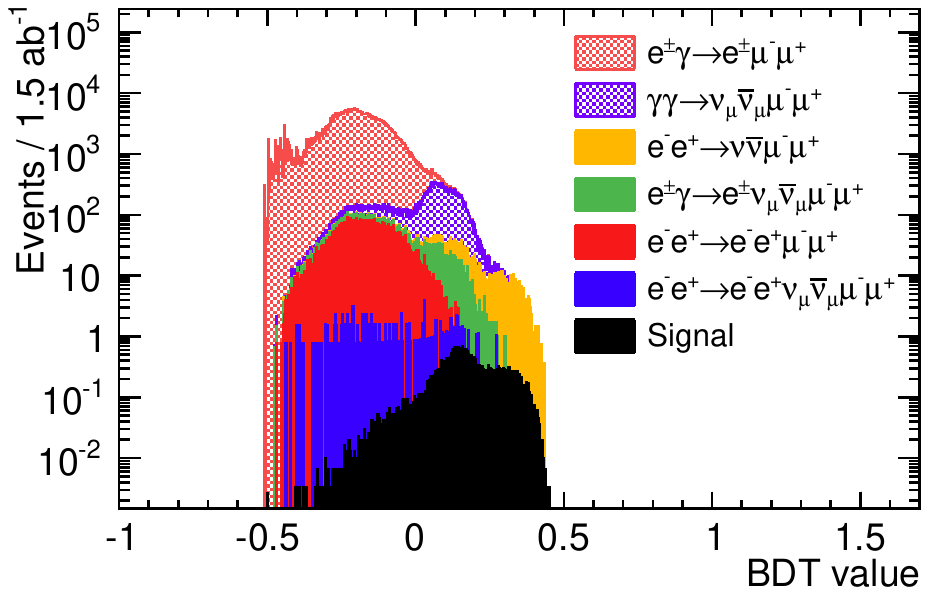}
\includegraphics[width=.4\textwidth]{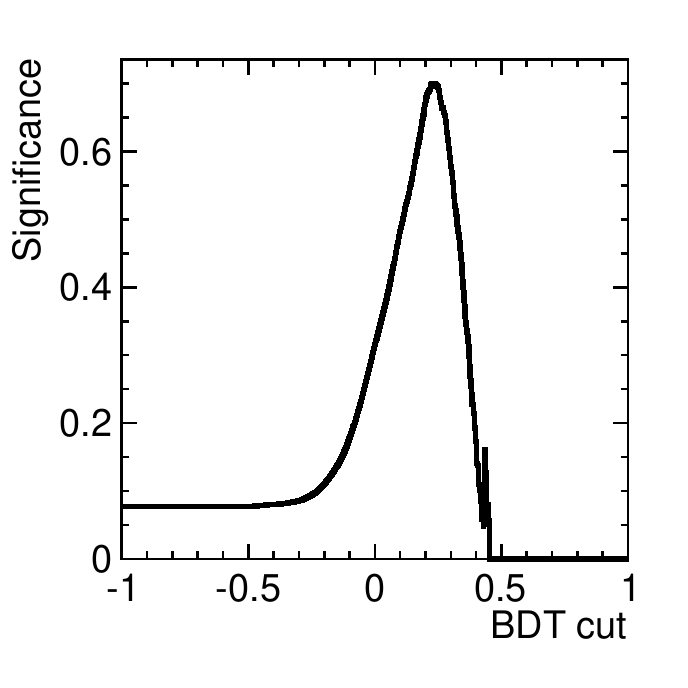}
\begin{textblock}{1.}(2.3,-5.9)
\bf CLICdp
\end{textblock}
\begin{textblock}{1.}(12.3,-5.9)
\bf CLICdp
\end{textblock}
\hspace*{.07\textwidth}(a)\hspace*{.53\textwidth}(b) 
\end{figure*}

\begin{figure*}
\centering
\caption{\label{fig-stackplot} Stacked histograms of the di-muon invariant mass distributions with preselecton only (a) and after MVA selection (b). The distributions are normalised to the same integrated luminosity of 
1.5 ab${^{-1}}$. The same legend applies to (a) and (b).}
\includegraphics[width=\textwidth]{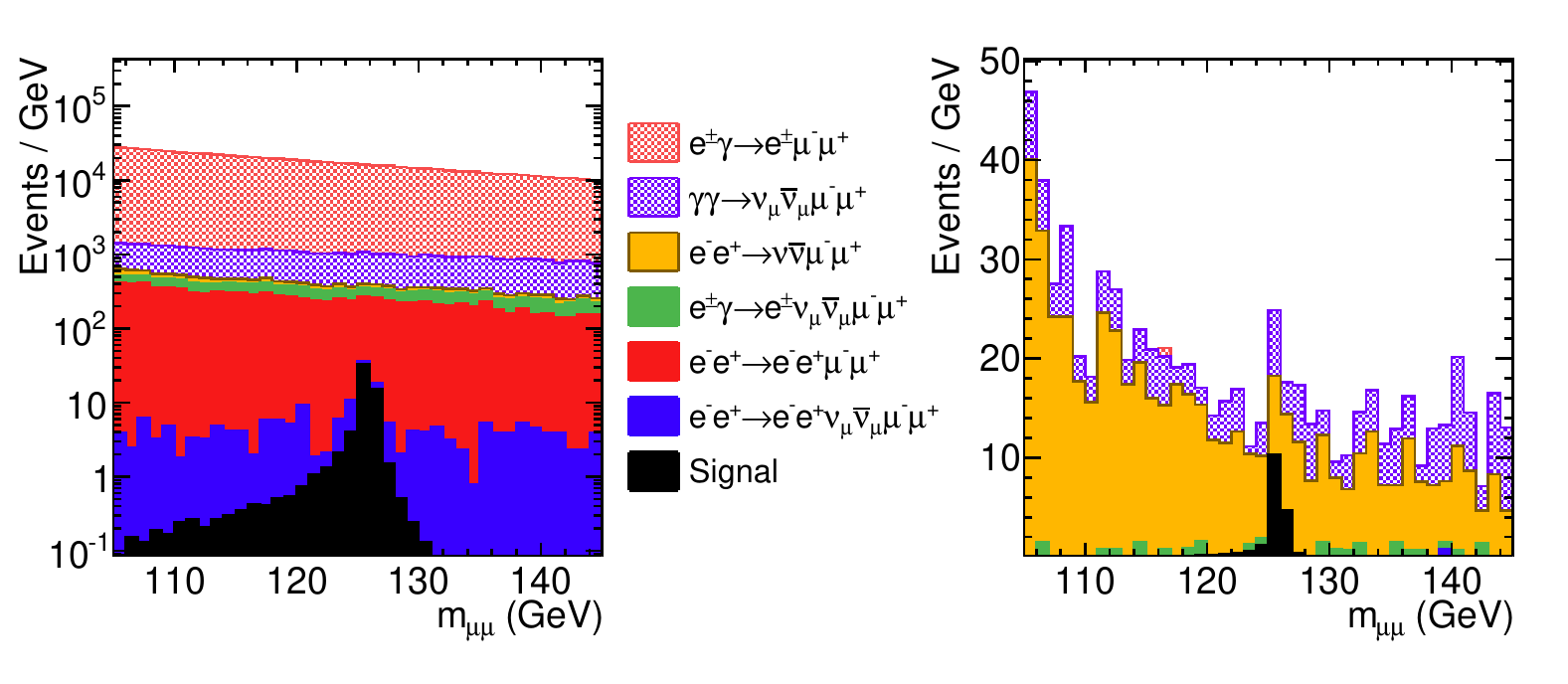}
\begin{textblock}{1.}(2.,-6.4)
\bf CLICdp
\end{textblock}
\begin{textblock}{1.}(12.3,-6.4)
\bf CLICdp
\end{textblock}
\hspace{.04\textwidth}(a)\hspace*{.53\textwidth}(b)
\end{figure*}

The distribution of the BDT classifier variable for the signal and the main background processes is shown in Figure \ref{fig-BDT} (a).
The classifier cut position was selected to maximise the significance, defined as $N_s / \sqrt{N_s + N_b}$, 
where $N_s$ and $N_b$ are the number of selected signal and background events, respectively. A plot of significance as a function of the position of the BDT cut is shown 
in Figure \ref{fig-BDT} (b). The optimal cut position was found at BDT = 0.23. Distributions of the di-muon invariant mass before and after the MVA selection are shown in Figure \ref{fig-stackplot}.
Figure \ref{fig-stackplot} (a) includes all events that pass the preselection, while Figure \ref{fig-stackplot} (b) shows all events passing the MVA selection. 
All samples are normalised to the integrated luminosity of 1.5 ab$^{-1}$. The signal preselection efficiency is 82\%.
The MVA selection efficiency for the signal is 32\%, reflecting the fact that sensitive observables have limited power to discriminate between the signal and background. The overall signal efficiency including reconstruction, preselection, losses due 
to coincident tagging of Bhabha particles and the MVA is 24\%, resulting in an expected number of 19 signal events.

\vspace{15mm}

\section{Di-muon invariant mass fit}\label{section7}

The quantity ${\sigma(H\nu\bar{\nu})\times BR(H\rightarrow\mu^+\mu^-)}$ is determined from the equation:
\begin{equation}
\label{eq-sigma}
	\sigma(H\nu\bar{\nu})\times BR(H\rightarrow\mu^+\mu^-) = \frac{N_s}{L \cdot \epsilon_s}
\end{equation}
where $L$ stands for the integrated luminosity and $\epsilon_s$ is the total counting efficiency for the signal, including the reconstruction, preselection and MVA selection. 
In the experiment, the number of signal events $N_s$ will be determined by fitting the di-muon invariant mass distribution with a function $f(m_{\mu\mu})$: 
\begin{equation}
\label{eq-s+b-model}
	f(m_{\mu\mu}) = N_s f_s(m_{\mu\mu}) + N_b f_b(m_{\mu\mu})
\end{equation}
where $f_{s,b}$ are probability density functions (PDF) used to describe the signal and the sum of all background processes, and $N_s$ and $N_b$ are the respective numbers of signal and background events in the 
fitting mass window. In this analysis, an unbinned likelihood fit, with all parameters of $f_{s,b}(m_{\mu\mu})$ fixed, is performed on simulated signal and background samples. $N_s$ and $N_b$ are left
as free parameters determined from the fit. The way the signal and background PDFs are obtained is discussed in Section \ref{section71}.

In order to estimate the statistical uncertainty of the signal count, 5000 toy Monte Carlo (MC) experiments are performed, where pseudo-data are obtained by randomly picking the signal $m_{\mu\mu}$ values
from the fully simulated signal sample, while background $m_{\mu\mu}$ values are randomly generated from the total background PDF $f_b(m_{\mu\mu})$. The size of the signal sample $N^\prime _s$ and sample sizes $N^\prime _{b,i}$ of individual backgrounds considered, 
are obtained from the Poisson distribution for the integrated luminosity of 
1.5 ab${^{-1}}$, taking into consideration corresponding cross-sections $\sigma$ and the selection efficiencies $\epsilon$ ($\left< N^\prime _{s}\right>=L\cdot\sigma_{s}\cdot\epsilon_{s}$, 
$\left< N^\prime_{b,i} \right>  = L \cdot \sigma_{i} \cdot \epsilon_{i}$, where i is indexing the different background processes listed in Table \ref{tableprocess}).

For each toy MC experiment, the $m_{\mu\mu}$ distribution is fitted by the function $f(m_{\mu\mu})$ 
given in Eq.\ref{eq-s+b-model}, and the standard deviation of the resulting distribution of $N_s$ over all toy MC experiments is taken as the estimate of 
the statistical uncertainty of the ${\sigma(H\nu\bar{\nu})\times BR(H\rightarrow\mu^+\mu^-)}$ measurement.

As will be discussed in Section \ref{section10}, the di-muon invariant mass distribution is 
sensitive to the detector $p_\text{T}$ resolution, while the Higgs width \ensuremath{\Gamma_{H}} can be considered negligible in comparison to the detector energy resolution.

\subsection{Signal and background PDFs}\label{section71}

Fully simulated samples of signal and background (Table \ref{tableprocess}) are fitted to extract the PDFs. The sizes of the samples vary from several tens of thousands of events for the signal, 
up to a few million of events for various background processes.

The signal PDF was defined as a linear combination of a Gaussian function with exponential tails, $f_{exp}$ and a Gaussian function with tails that asymptotically approach constant values in the high and low $m_{\mu\mu}$, $f_{flat}$:
\begin{eqnarray}\label{eq3}
 f_s &=& f_{flat} + C \cdot f_{exp} \nonumber \hspace{35pt} \text{where}
\\
f_{flat} &=& \left\{ \begin{array}{rl}
		e^{-\frac{ (m_{\mu\mu} - m_{H})^2 }
          {2 \sigma^2 + \beta_L (m_{\mu\mu} - m_{H})^2 } }
	&	m_{\mu\mu} < m_{H} \\
		e^{-\frac{ (m_{\mu\mu} - m_{H})^2 }
          {2 \sigma^2 + \beta_R (m_{\mu\mu} - m_{H})^2 } } 
	& 	m_{\mu\mu} > m_{H} 
	\end{array} \right. \label{eq-pdf-signal}  \hspace{25pt}   \text{and}\\
f_{exp} &=& \left\{ \begin{array}{rl}                              
		e^{-\frac{ (m_{\mu\mu} - m_{H})^2 }
          {2 \sigma^2 + \alpha_L |m_{\mu\mu} - m_{H}| } }
	&	m_{\mu\mu} < m_{H} \\
		e^{-\frac{ (m_{\mu\mu} - m_{H})^2 }        
          {2 \sigma^2 + \alpha_R |m_{\mu\mu} - m_{H}| } } 
	&	m_{\mu\mu} > m_{H} \hspace{10pt} .
	\end{array} \right . \nonumber  
\end{eqnarray}

\begin{figure}[H]
\centering
\caption{\label{fig-SignalFit}Distribution of the di-muon invariant mass $m_{\mu\mu}$, of the signal after MVA selection, used for the PDF determination. The distribution is fitted with the function $f_s$ given in Eq.\ref{eq3}.}
\includegraphics[width=\columnwidth]{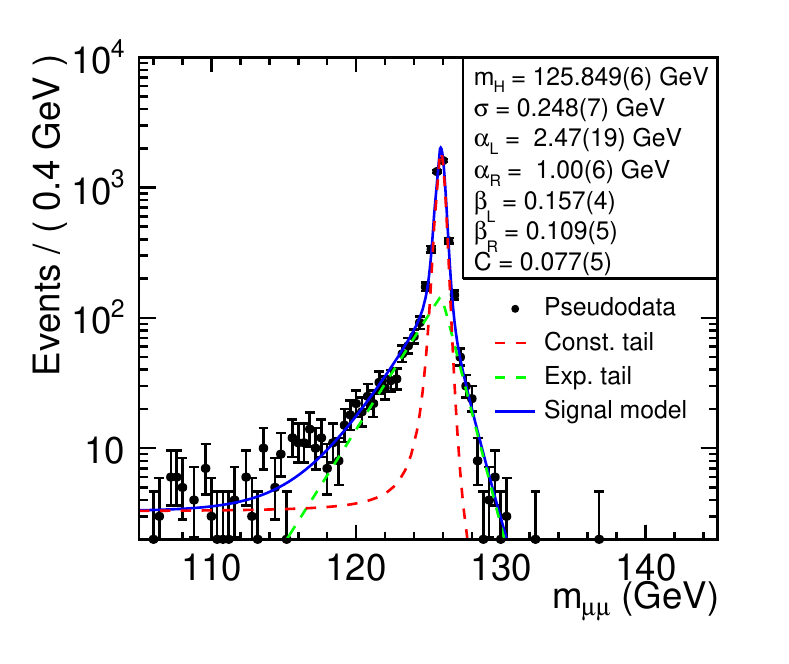}
\begin{textblock}{1.}(2.,-5.9)
\bf CLICdp
\end{textblock}
\end{figure}

The parameters of Eq. \ref{eq3} are determined by fitting the di-muon invariant mass distribution for the signal (Figure \ref{fig-SignalFit}). 

The total background PDF is defined as a linear combination of a constant and exponential term:
\begin{equation}
f_b = p_0\cdot(p_1 e^{p_2(m - m_{H})} + (1 - p_1))
\label{eq-bkg-pdf}
\end{equation}

\begin{figure}
\centering
\caption{\label{fig-bkgFit}Distribution of the di-muon invariant mass $m_{\mu\mu}$, of the total background, after MVA selection. The distribution is fitted with the function $f_b$ given in Eq. \ref{eq-bkg-pdf}}
\includegraphics[width=\columnwidth]{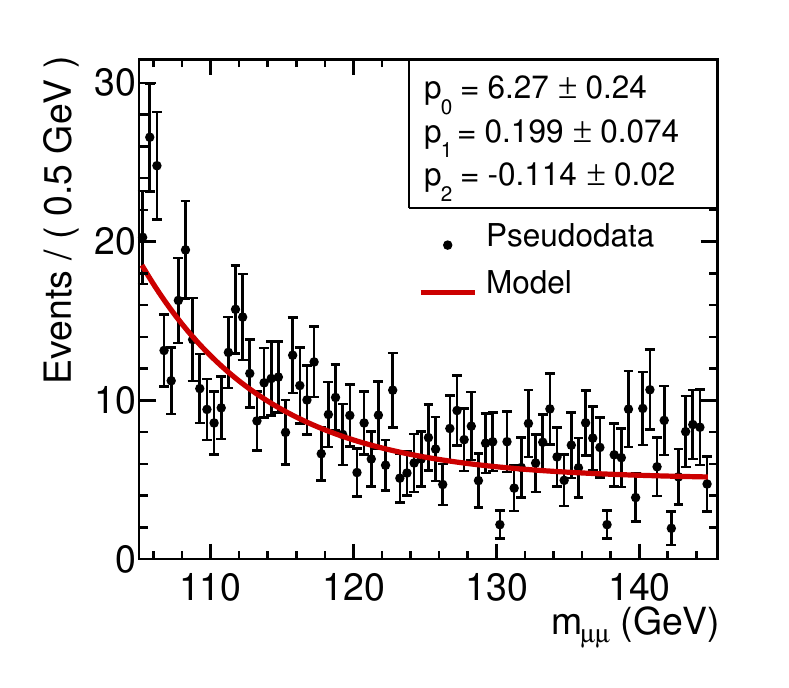}
\begin{textblock}{1.}(2.,-5.9)
\bf CLICdp
\end{textblock}
\end{figure}

The di-muon invariant mass fit of the total background is shown in Figure \ref{fig-bkgFit}, together with the fit results for the free parameters in
Eq.\ref{eq-bkg-pdf}. As the normalisation to the common integrated luminosity requires different normalisation coefficients for different processes, 
binned data were used to combine the background processes in a straightforward manner and a binned $\chi^2$ fit was performed. The $\chi^2 / N_{df}$ of the background fit was 62/61.  

\subsection{Distribution of the signal count}\label{section72}

\begin{figure}
\centering
\caption{\label{fig-TMCfit}Fitted distribution of the di-muon invariant mass $m_{\mu\mu}$, for the sum of the signal and the total background in one toy MC experiment.}
\includegraphics[width=\columnwidth]{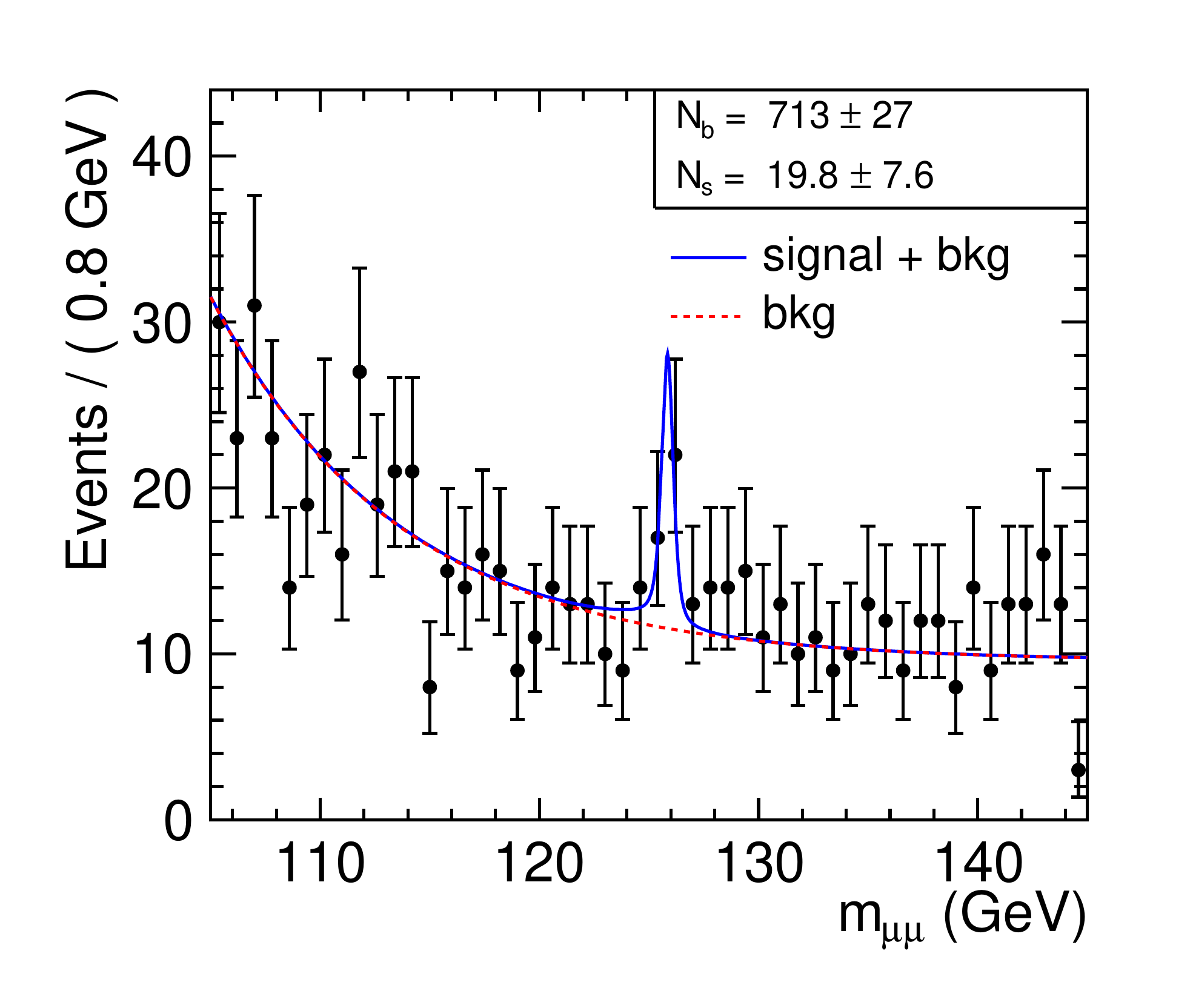}
\begin{textblock}{1.}(2.,-5.9)
\bf CLICdp
\end{textblock}
\end{figure}

The overall function  $f(m_{\mu\mu})$ (Eq.\ref{eq-s+b-model}) is fitted to the pseudo-data of each toy MC experiment using the unbinned likelihood fit. 
An example of a toy MC fit is given in Figure \ref{fig-TMCfit}.

\begin{figure}
\centering
\caption{\label{fig-distri} Distribution of the number of signal events in 5000 toy MC experiments.}
\includegraphics[width=\columnwidth]{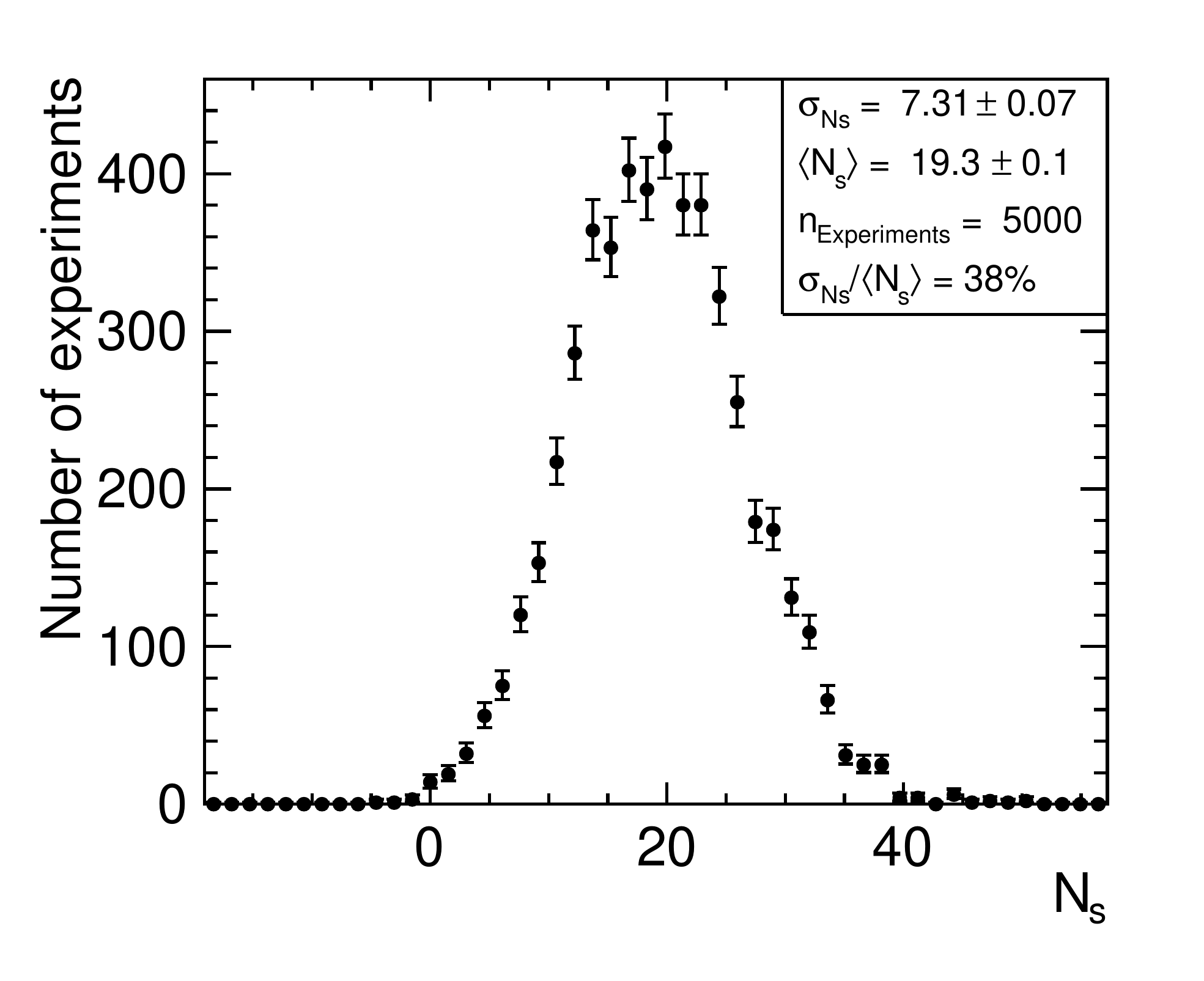}
\begin{textblock}{1.}(2.,-5.9)
\bf CLICdp
\end{textblock}
\end{figure}

The standard deviation of the resulting signal count distribution in 5000 repeated toy MC experiments corresponds to the statistical uncertainty of the measurement and is 38\%. 
(Figure \ref{fig-distri}). According to Eq.\ref{eq-sigma}  it translates into the statistical uncertainty of the ${\sigma(H\nu\bar{\nu})\times BR(H\rightarrow\mu^+\mu^-)}$ measurement, 
having in mind that the total uncertainty of the integrated luminosity can be determined at the permille level \cite{r34}.

The statistical uncertainty of the signal counting is dominated by contributions from the limited signal statistics and from a presence of irreducible backgrounds. 
To estimate the significance of the signal against the null-hypothesis, another set of 5000 toy MC experiments was performed with zero signal count, and $f(m_{\mu\mu})$ (Eq.\ref{eq-s+b-model})
was fitted to the pseudo-data. 
The resulting $N_s$ distribution was centered on zero with a standard deviation of 5.4. Thus, in the case where the SM expected number of 19 signal events are found in an experiment, 
the corresponding signal significance would be $3.7\,\sigma$.

The Higgs coupling to muons, $g_{H{\mu\mu}}$, is optimally extracted in a global fit procedure taking into account all Higgs measurements at the 350 GeV, 1.4 TeV and 3 TeV stages. 
The global fit serves to extract Higgs couplings from all measurements, as well as the experimental Higgs width ${\ensuremath{\Gamma_{H}}}$. 
Because ${\sigma(H\nu\bar{\nu})\times BR(H\rightarrow}$ ${\mu^+\mu^-)} \propto \frac{{g^2_{HWW} g^2_{H{\mu\mu}}}}{\ensuremath{\Gamma_{H}}}$ and having access to ${\ensuremath{\Gamma_{H}}}$ and $g_{HWW}$ from other measurements, extraction of $g_{H{\mu\mu}}$ is possible solely from the measurement presented here.
An example of a minimal set of measurements 
giving a model-independent access to ${\ensuremath{\Gamma_{H}}}$ and $g_{HWW}$ is the following: the  $H\rightarrow b\bar{b}$ measurements\thinspace at both 350 GeV and 1.4 TeV give access\thinspace to\thinspace the\thinspace ratio\thinspace 
$\frac{g_{HWW}}{g_{HZZ}}$, the recoil mass measurement at 350 GeV CM energy gives access to $g_{HZZ}$, and the $H \rightarrow W^+W^-$ measurement at 1.4 TeV gives access to the ratio 
$\frac{g^4_{HWW}}{\ensuremath{\Gamma_{H}}}$ \cite{r4}. The contributions of these measurements towards the final $\Delta g_{H{\mu\mu}}$ is negligible at the second significant digit.

The dominant contribution to the $g_{H{\mu\mu}}$ coupling uncertainty is the statistical uncertainty of the measurement presented here. 
Systematic uncertainties affect the total uncertainty of $g_{H{\mu\mu}}$ determination only at the third significant digit, and thus can be neglected (Section \ref{section9}). 
Under these assumptions, the relative uncertainty of $g_{H{\mu\mu}}$ is approximated to be 19\%.

\section{Impact of electron polarization}\label{section8}

If 80\% left-handed polarisation of the electron beam is assumed during the entire operation time at 1.4 TeV, the Higgs production cross-section through $WW$-fusion would be enhanced by a factor 1.8 \cite{r4}. 
The most important background contribution after the MVA selection, the $e^+e^-\rightarrow\nu_{e}\bar{\nu_{e}}\mu^+\mu^-$ process, 
is enhanced by the same factor because it is also mediated by $W$ bosons which have only left-handed interactions. The process ${e^\pm}\gamma\rightarrow{e^\pm}\mu^+\mu^-$ is enhanced by a factor 1.32, 
while cross-sections for other background processes are not significantly changed w.r.t. the unpolarised case. The overall selection efficiency of the signal is 30\%, 
because the classifier cut position is moved to a lower value which consequently leads to a higher signal efficiency. The final statistical uncertainty of the ${\sigma(H\nu\bar{\nu})\times BR(H\rightarrow\mu^+\mu^-)}$ 
measurement is 25\%.
The corresponding uncertainty of $g_{H{\mu\mu}}$ is 13\%.
A summary of the results of the presented analysis is given in Table \ref{tab-results}. It is important to note that all kinematic variables are unaffected by the beam polarization.

\begin{table}
\centering
\caption{\label{tab-results}Summary of the ${\sigma(H\nu\bar{\nu})\times BR(H\rightarrow\mu^+\mu^-)}$ measurement at 1.4 TeV CLIC with unpolarised and 80\% polarized electron beams. All uncertainties are statistical.}
\begin{tabular*}{\columnwidth}{@{\extracolsep{\fill}}lll@{}}
\hline
\multicolumn{1}{c}{} & \multicolumn{1}{c}{Unpolarised}  & \multicolumn{1}{c}{Polarised (80\%, 0\%)}\\
\hline
 $N_s$ & $19.3 \pm 0.1$  & $35 \pm 9$ \\
 $\epsilon_s$   & 24\%  & 25\%\\
 $\frac{\delta(\sigma(H\nu\bar{\nu})\times BR(H\rightarrow\mu^+\mu^-) )}{\sigma(H\nu\bar{\nu})\times BR(H\rightarrow\mu^+\mu^-)}$ & 38\% & 25\%\\
 $\delta(g_{H_{\mu\mu}})/g_{H_{\mu\mu}}$      & 19\%  & 13\% \\
\hline
\end{tabular*}
\end{table}

\section{Systematic uncertainties}\label{section9}

From Eq.\ref{eq-sigma} it is clear that uncertainties of the integrated luminosity and muon identification efficiency influence the uncertainty of the $H\rightarrow\mu^+\mu^-$ branching ratio measurement at 
the systematic level. It has been shown that at 3 TeV CLIC \cite{r35}, where the impact of the beam-induced processes is the most severe, the luminosity above 75\% of the nominal CM energy 
can be determined at the permille level, using low-angle Bhabha scattering. Below 75\% of the nominal CM energy, the luminosity spectrum can be measured with a precision of a few percent using wide-angle Bhabha scattering \cite{r36}. 
About 17\% of all Higgs production events occur at a CM energy below 75\% of the nominal CM energy. Having in mind the intrinsic statistical limitations of the signal sample, this source of systematic uncertainty can be considered negligible.

\begin{figure}
\centering
\caption{\label{fig-bias} Impact of the uncertainty of the muon $p_\text{T}$ resolution on the signal counting. The relative shift of the signal count is given as a function of the relative shift of the $p_\text{T}$ resolution. }
\includegraphics[width=\columnwidth]{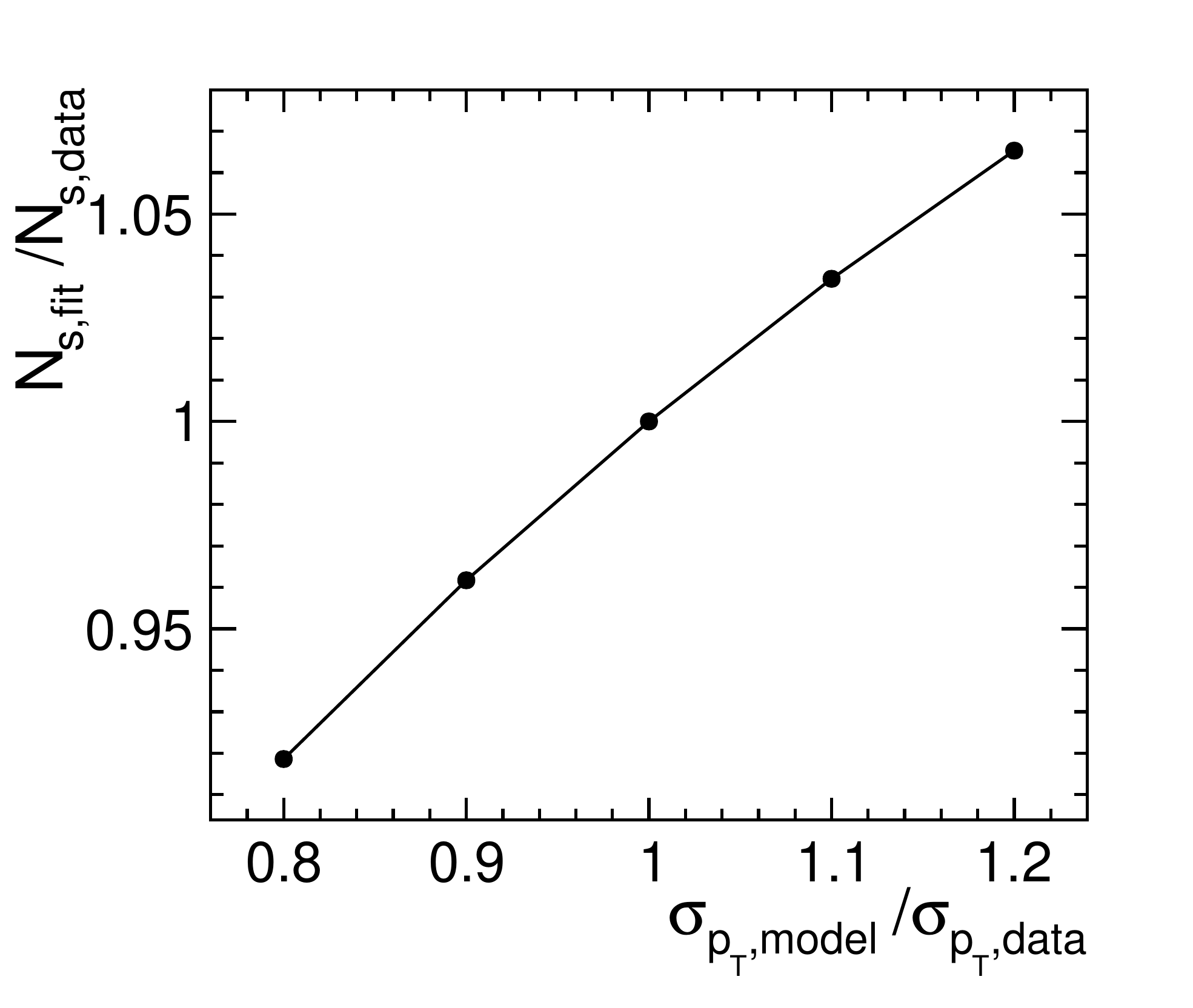}
\begin{textblock}{1.}(2.,-5.9)
\bf CLICdp
\end{textblock}
\end{figure}

On the detector side, an important systematic effect is the uncertainty on the transverse momentum resolution, because it directly influences the expected shape of the signal $m_{\mu\mu}$ distribution. The sensitivity of the signal count to the accuracy of the knowledge of the $p_\text{T}$ resolution $\sigma_{p_\text{T}}$  has been studied by performing the analysis with an artificially introduced uncertainty of an exaggerated magnitude on the assumed ${p_\text{T}}$ resolution used to extract the signal PDF. 
Results of the relative shift in signal counts w.r.t. the relative shift of $\sigma_{p_\text{T}}$ are shown in Figure \ref{fig-bias}. 
The relative bias in signal counting per one percent change of $\sigma_{p_\text{T}}$ is 0.35\%.

The uncertainty of the muon identification efficiency will directly influence the signal selection efficiency. In addition, the uncertainty of the muon polar angle resolution impacts the $m_{\mu\mu}$ reconstruction. Based on the results of the LEP experiments \cite{r37}, it can be assumed that these detector related uncertainties are below a percent.

The systematic uncertainty of the signal count caused by the fit with $f_{m_{\mu\mu}}$ defined in Eq.\ref{eq-s+b-model}, was found to be about 1\% which is small compared to the statistical error.

Because of the forward EM shower tagging, about 7\% of all events are rejected by coincident detection of Bhabha events.
This fraction must be precisely calculated taking into account Bhabha event distributions, beam-beam effects,
as well as the dependence of the tagging efficiency on energy and angle of the incident electrons and photons.
This is work in progress \cite{r38, r39, r40}, but the uncertainty of this effect is also expected to be negligible compared to the statistical uncertainty of the measurement. 

\subsection{Benefit of a improved $p_\text{T}$ resolution}\label{section10}

\begin{figure}
\centering
\caption{\label{fig-pTres} Dependence of the relative statistical uncertainty of the ${\sigma(H\nu\bar{\nu})\times BR(H\rightarrow\mu^+\mu^-)}$ on the transverse momentum resolution, $\delta_{1/p_\text{T}}$, averaged over the signal sample in the whole detector.}
\includegraphics[width=\columnwidth]{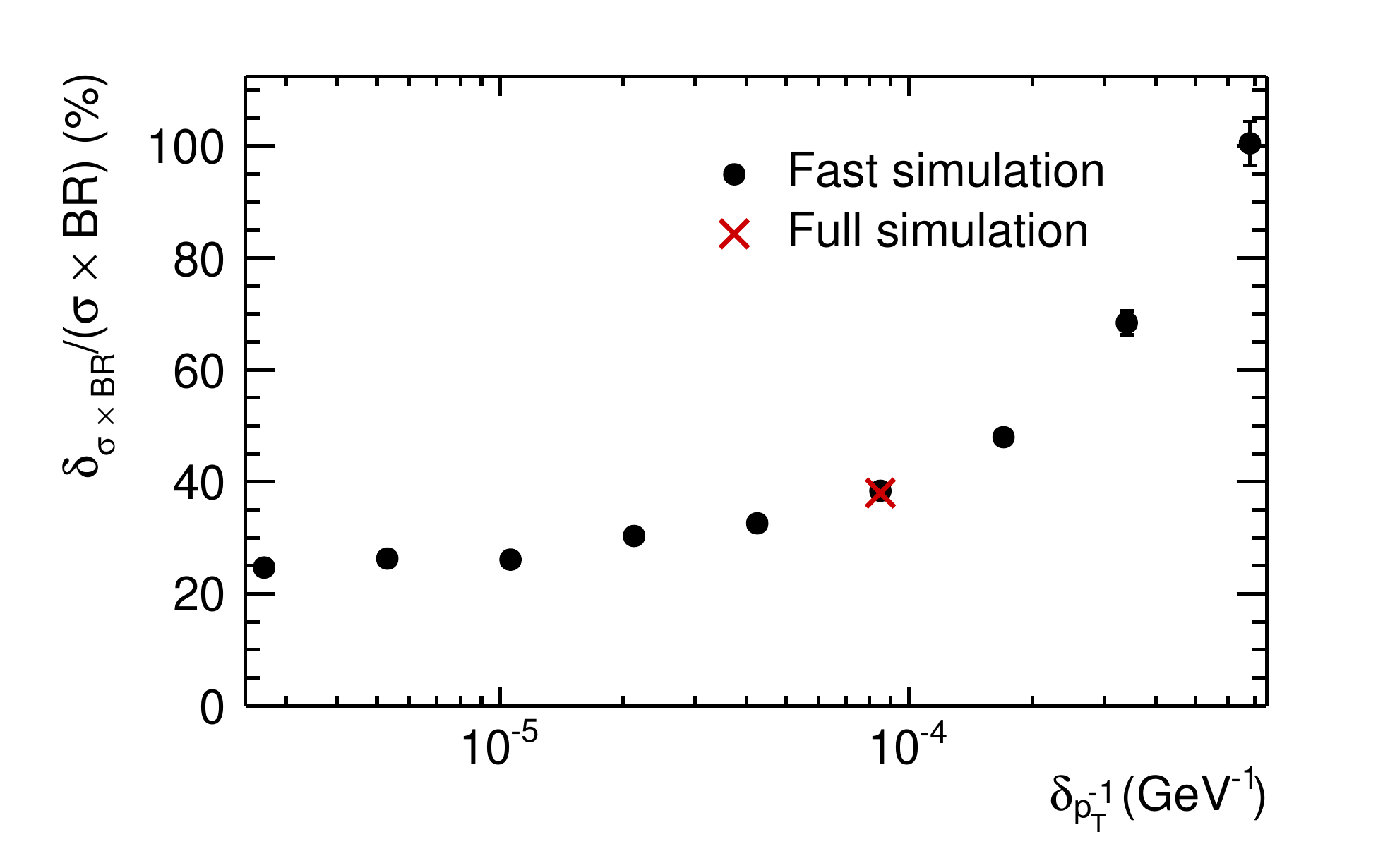}
\begin{textblock}{1.}(2.,-4.5)
\bf CLICdp
\end{textblock}
\end{figure}

To estimate the benefit of a better $p_\text{T}$ resolution, the analysis was repeated by substituting the muon four-momenta reconstructed in the full simulation of the 
signal by the four-momenta obtained by a parametrisation of the momentum resolution for several different values of the detector resolution. Figure \ref{fig-pTres} 
displays the approximate dependence of the statistical uncertainty of the measurement on the average transverse momentum resolution in the whole detector. 
Even a large improvement of the muon momentum resolution would result in only a moderate improvement of the statistical uncertainty of 
the measured product of the Higgs production cross-section and the branching ratio for the $H\rightarrow\mu^+\mu^-$ decay. 

\section{Conclusions}
\label{section11}
It has been shown that the measurement of the cross-section times the branching ratio for the SM Higgs decay into two muons can be performed
with a relative statistical uncertainty of 38\% at 1.4 TeV
CLIC, assuming 1.5 ab$^{-1}$ integrated luminosity with unpolarised beams. The result
is dominated by the limited signal statistics and the
irreducible background. The systematic uncertainties are negligible in comparison to the statistical one. This translates into a relative
uncertainty of the coupling of Higgs to
muons $g_{H{\mu\mu}}$ of aproximately 19\%.
If the same integrated luminosity is collected with 80\% left-handed polarisation for the electrons,
the relative statistical uncertainty improves to 25\% and 13\% for ${\sigma(H\nu\bar{\nu})\times BR(H\rightarrow\mu^+\mu^-)}$ and $g_{H{\mu\mu}}$, respectively.

\begin{acknowledgements}
The authors would like to thank the members of the analysis working 
group in the CLICdp collaboration for useful discussions. Konrad 
Elsener, Aharon Levy, Sophie Redford and Eva Sicking are especially 
acknowledged for carefully reading the manuscript. The production of the 
investigated event samples would not have been possible without the 
support from Stephane Poss and Andre Sailer. We acknowledge the support received from the Ministry of education, science and technological development of the Republic of Serbia within the project OI171012.
\end{acknowledgements}

\end{document}